\documentclass[10pt]{article}

\usepackage{bsymb,b2latex}
\usepackage{ifxetex}
\usepackage{mathpazo}
\usepackage{graphicx}
\usepackage{float}
\usepackage[a4paper,top=3cm, bottom=2.5cm, left=2.5cm, right=2.5cm,
headsep=1.25cm,headheight=34pt]{geometry}
\ifxetex
\usepackage{fontspec}
\usepackage{xltxtra}
\else
\usepackage[T1]{fontenc}
\usepackage[utf8]{inputenc}
\fi
\usepackage[colorlinks=true,urlcolor=black,linkcolor=black,citecolor=black]{hyperref}

\title{Formal Data Validation with Event-B}

\author{
  Frédéric Badeau\\ Systerel \\1090 rue René Descartes\\Parc
  d'activités de la Duranne\\13857 Aix-en-Provence cedex 3\\
  frederic.badeau@systerel.fr
\and
  Marielle Doche-Petit\\ Systerel \\1090 rue René Descartes\\Parc
  d'activités de la Duranne\\13857 Aix-en-Provence cedex 3\\
  marielle.doche-petit@systerel.fr
       }

\begin{document}
\maketitle

\abstract{This article presents a verification and validation activity performed in an industrial context, to validate configuration data of a metro CBTC system by creating a formal B model of these configuration data and of their properties. A double tool chain is used to safely check whether a certain given input of configuration data fulfill its properties. One tool is based on some Rodin and open source plug-ins and the other tool is based on ProB.}

\section{Data configuration of CBTC}
A Communication Based Train Control (CBTC) is a system used to safely control metro systems. It allows several train control modes including a fully automatic mode. It must achieve high safety-critical and high availability levels defined in the CENELEC standards (EN 50126 \cite{En50126}, EN 50128
\cite{En50128}, EN 50129 \cite{En50129}).

A CBTC system comes with a lot of configuration data because these data should describe a large part of the metro railway network. Much equipment of the CBTC should be parametrized with these configuration data. Each piece of equipment typically uses from a few kilo-bytes to one or more mega-bytes of data.

Pieces of CBTC equipment are controlled by software parametrized by configuration data. Although the process presented here could be applied on both safety-critical and non safety-critical parts, it is only applied on safety-critical parts in order to reduce development costs. Safety-critical software is developed independently from its configuration data, thus a software version has to be validated only once, even if it is instantiated for several pieces of equipment. It also does not have to be validated again after some configuration data change. The data properties as required by a piece of software, are described as requirements in the software interface document. Every set of configuration data also has to be validated once.

For the CBTC system under development, this principle of separation between software and configuration data is taken to the extreme, since configuration data are not built-in into equipment, they are instead dynamically loaded at runtime through configuration messages.

Although configuration data are just built to parametrize some equipment, these data may be partial: each piece of equipment needs only access to its own specific data subset, otherwise some data can be shared beetween several pieces of equipment. 
So the development process of configuration data starts by developing a common database of configuration data representing the whole CBTC system. This database should have no or very low redundancy. 

\section{Validation of configuration data}

To validate configuration data we build a B model of the configuration data and of their properties with basic B types and predicates. Then, given an input set of configuration data we use a double tool chain to evaluate all the predicates for these configuration data.

The process of validating configuration data is split into two major activities:
\begin{itemize}
\item the preparation phase which consists in modelling data and their properties and then verifying and testing the model,
\item the actual validation of a given configuration data set against the obtained model.
\end{itemize}

The preparation activity takes by far the most time, whereas the actual validation is more about setting everything up and then running automated tools.

\begin{figure}
\begin{center}\label{process}
  \includegraphics[width=4in]{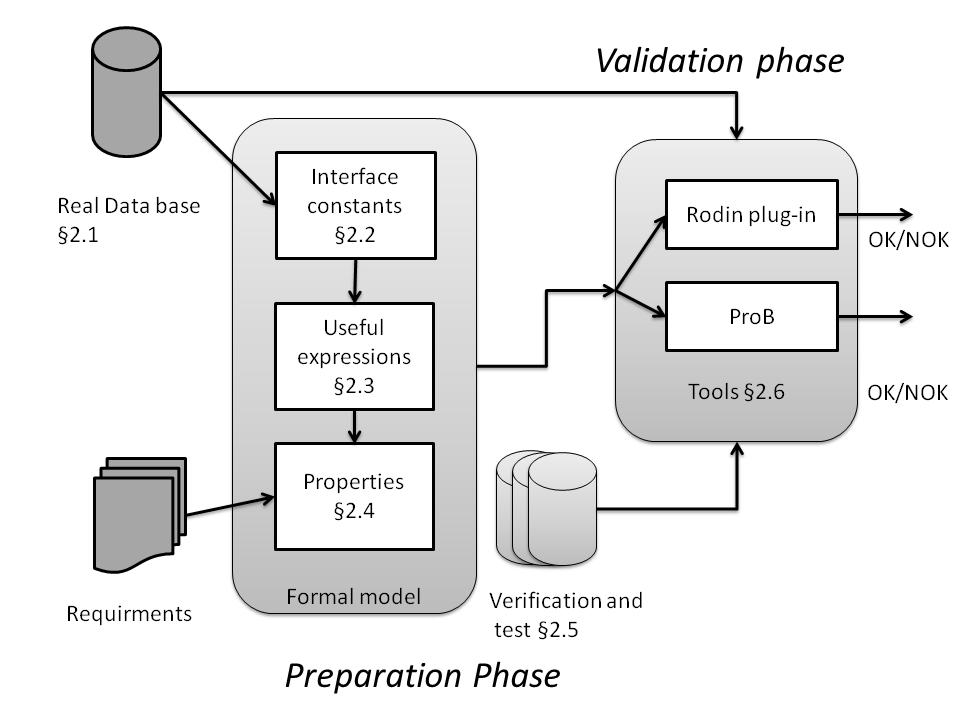}
\end{center}
\caption{Validation process}
\end{figure}

\subsection{The real database}

In one case, data are given in the XML format derived from a UML model. XML is well-suited to be interfaced with the predicates evaluation tools.\footnote{Our approach can be adapted to other classical database interfaces like SQL, excel,...}  

However, it is necessary to model and access data from binary messages derived from the XML database, because these messages contain the actual configuration data that are loaded into safety-critical equipment of the CBTC system.

Interfacing predicate evaluation tools directly with binary structures is not recommended, since it would depend too much on a particular binary structure and would be difficult to maintain. It has been decided to develop a tool to convert a binary structure into an XML file. To deal with safety issues, a reverse tool has also been independently developed to convert back the XML file into a binary file. To ensure that the binary/XML converter works fine, we apply the reverse XML/binary converter and check that the result is the same as the binary file we started from. 

\subsection{Modelling interface constants}
The preparation activity starts by interfacing the data of the real database with B constants, which will provide the basic bricks of the formal model.

The following types\footnote{Usually two-dimensional functions are enough to model everything we need from the real configuration database.} of constants can be declared as basic interface constants to link elements of the real database to constants of the B properties:
\begin{itemize}
\item carrier sets,
\item subsets of carrier sets,
\item scalar data (mostly integers),
\item functions from a scalar type to another scalar type,
\item relations between a scalar type and another scalar type,
\item functions from a scalar type to functions from a scalar type into another scalar type.
\end{itemize}

For all these cases, a constant is given by its name, a typing predicate in the B language (except for carrier sets as they define new scalar types) and an XPath request. When these requests are placed on some XML file, they select one value, or a set of values. In the case of functions and relations, several XPath requests are required, to select first a domain element, and then to select a range element. And in the case of a function of functions three XPath requests are required, two for the domain and one for the range.

Carrier sets are used to define every distinct type of objects handled in the database, such as lines, sections, blocks, block frontiers, track circuits, switches, signals (traffic lights), beacons, etc. In a CBTC system, a block is an elementary linear part of a railroad which defines a local one dimension coordinate system oriented by its two frontiers. By definition, the origin frontier is located at abscissa 0 and the destination frontier is located at abscissa block length.

\begin{center}
  \begin{minipage}{.5\textwidth}
    \begin{description}
	\SETS
	\begin{description}
	    \Item{ t\_block; }
	    \Item{ t\_block\_frontier }
	\end{description}

    \end{description}
  \end{minipage}
\end{center}

Scalar data are used to define constant numbers of the system, such as special distances, speeds, accelerations, delays.

Functions and relations are used to define links between objects, such as the length of a block, the block a signal is located on, the signal abscissa, the blocks that follow a given block in a given direction. Functions and relations are the most common building units handled by the model.

The following interface constants represent the block frontier located at the origin or at the destination of a block. Destination and origin define the block orientation.

  \begin{center}
    \begin{minipage}{.6\textwidth}
      \begin{description}
	  \Item{ f\_block\_orig \in t\_block \tfun t\_block\_frontier \land }
	  \Item{ f\_block\_dest \in t\_block \tfun t\_block\_frontier}
      \end{description}
    \end{minipage}
  \end{center}
   
These interface constants do not have to be a straightforward copy of the real database that could be produced automatically. They are the first step of the model and should be the result of modelling choices in order to make the model as simple and as effective as possible to express properties. For instance, in the XML database a signal has an enumerated attribute to state its type between manoeuvre signal, spacing signal, or permanent signal. Keeping this attribute is something to do at a programming language level. In a set-theoretic model, we prefer to define constant subsets for each type of signal.

Concerning floating point values, we reach a limitation of the B Language, which only supports integers, so we use only fixed point integers isomorphic to integers in trivial way. The XML database contains floating point values. The conversion to integers can be done with XPath operators.

\subsection{Modelling useful expressions}

The second modelling step consists in building a library of useful formal expressions which will serve as building blocks for the whole model.
These intermediate useful expressions are called definitions, they are similar to a \texttt{LET} in software-B. A definition is associated with a name, an informal description, and a B expression.

Although interface constants provide a certain flexibility, they may be too limited to define directly the data structures best suited to model properties. For instance, interface constant functions are limited to two-dimensional functions.
Definitions based on interface constants are used to define the constants we really want for the B model.

Thanks to the definitions, the B model is easier to read, to understand and to verify. This breaking down mechanism into many intermediate steps does really work fine and does not come with any downside, so it may be used intensively. Especially, there is no limitation on the number of nested definitions used in a B model and the tools do not have any specific time efficiency issues due to nested definitions.

For the CBTC under development, a library of definitions was used to model graph functions. Indeed, a CBTC railway network is represented by an oriented graph of blocks. The following properties hold on blocks:
\begin{itemize}
  \item A block has two frontiers noted \emph{up} and \emph{down}.
  \item The down frontier is located at the origin of the block at abscissa 0 and the up frontier is located at the destination of the block at abscissa \emph{block length}.
  \item A block frontier may be connected to no other block, to one block, or to two blocks if there is a switch at this block frontier and if the block is at the narrow end of the switch. In the latter case, the two following blocks are called left block and right block depending on their position for an observer located on the narrow end and facing the switch point.
\end{itemize}

The four constants below represent the possible next block connected to a block depending on the direction (upward/downward) and the possible turning direction (left/right).

  \begin{center}
    \begin{minipage}{.8\textwidth}
      \begin{description}
	  \Item{f\_next\_upward\_left\_block    \in t\_block \pfun t\_block \land}
	  \Item{f\_next\_downward\_right\_block \in t\_block \pfun t\_block \land}
	  \Item{f\_next\_upward\_right\_block   \in t\_block \pfun t\_block \land}
	  \Item{f\_next\_downward\_left\_block  \in t\_block \pfun t\_block}
      \end{description}
    \end{minipage}
  \end{center}

These functions are partial functions which allows to take account of the case where there is no following block (a convention then fixes the direction which gives the next block).

We define the relation of the possible next block in a direction (upward/downward) this way:

  \begin{center}  \begin{minipage}{\textwidth}
    \begin{description}
	\small
	\Item{r\_next\_upward\_block   = f\_next\_upward\_left\_block   \cup  f\_next\_upward\_right\_block      \land}
	\Item{r\_next\_downward\_block = f\_next\_downward\_left\_block \cup   f\_next\_downward\_right\_block}
    \end{description}
  \end{minipage}
  \end{center}

The railway graph is then modelled by a relation associating an oriented block with every oriented block it is connected to. The fact that block orientation is arbitrary, meaning that a block oriented in a direction may be followed in this direction by a block oriented in the opposite direction makes the model more complex. However this is required to model correctly all possible railway network topologies.

Then we define the relation that associates to a block and a direction, each next block in the corresponding direction:

\begin{center}
 \begin{description}
	\small
	\Item{r\_next\_block = }
	\Item{((r\_next\_upward\_block \cap (f\_block\_dest   \fcomp f\_block\_origin^{-1})) \pprod \{c\_upward \mapsto c\_upward\}) \cup }
	\Item{((r\_next\_upward\_block \cap (f\_block\_dest   \fcomp f\_block\_dest^{-1})) \pprod \{c\_upward \mapsto c\_downward\}) \cup }
	\Item{((r\_next\_upward\_block \cap (f\_block\_origin \fcomp f\_block\_dest^{-1})) \pprod \{c\_downward \mapsto c\_downward\}) \cup }
	\Item{((r\_next\_upward\_block \cap (f\_block\_origin \fcomp f\_block\_origin^{-1})) \pprod \{c\_downward \mapsto c\_upward\}) }
    \end{description}

\end{center}   

Then, we build a library of functions dealing with this oriented graph. Some ordering functions define whether a position given by an abscissa on a block is located after another given position, with respects to a certain direction. They all use the iterate operator on the relation graph. Several ordering functions are defined depending on which block direction is used as the reference direction.

We define the relation that associates to a block and a direction, every downstream blocks in the corresponding direction:

  \begin{center}
    \begin{minipage}{.5\textwidth}
      \begin{description}
	  \Item{r\_block\_chain = closure1(r\_next\_block)}
      \end{description}
    \end{minipage}
  \end{center}

The following function states whether a position 2 is located afterwards a position 1 in a given direction related to the orientation on block 1.

  \begin{center}
    \begin{minipage}{.8\textwidth}
      \small
      \begin{description}
	  \Item{f\_pos\_afterwards = }
	  \Item{\% dir1\mapsto block1\mapsto abs1
	                    \mapsto block2\mapsto abs2\qdot}
	  \begin{description}
	      \Item{    (dir1   \in t\_dir   \land }
	      \Item{    block1 \in t\_block \land }
	      \Item{   abs1   \in INTEGER \land }
	      \Item{    block2 \in t\_block \land }
	      \Item{    abs2   \in INTEGER | }
	      \Item{    \bool( }
	      \begin{description}
		  \Item{	 (block1 = block2 }
		  \Item{         \limp }
		  \Item{         \;\;\{c\_downward \mapsto \bool(abs2 \leq abs1), }
		  \Item{	 \;\;\; c\_upward   \mapsto \bool(abs1 \leq abs2)\}(dir1) = \True) \land }
		  \Item{	 (block1 \neq block2 }
		  \Item{         \limp }
		  \Item{        \;\;\exists dir2\qdot(}
			  \begin{description}
				  \Item{ dir2 \in t\_dir \land}
			  	  \Item{(block1 \mapsto dir1) \mapsto (block2 \mapsto dir2) \in r\_block\_chain)))) }
		          \end{description}
	      \end{description}
	  \end{description}
      \end{description}
    \end{minipage}
  \end{center}

Many properties deal with distances on the railway graph. Although these properties seem straightforward in natural language, they are actually more complex than expected to handle, because there could be several paths between two positions. To handle distances, we define a zone, which is a graph sub-part, and we define library functions giving the zone obtained by starting from a direction and moving on to a certain maximum distance. Several functions are defined to suit all the properties dealing with distances.

\subsection{Modelling properties}

Every property requirement on configuration data coming from the relevant documents produced during system design should be modelled by a predicate, which is the last step of creating the B model. Every property is given a name, the design requirement it refers to, a predicate and a natural language description of the predicate. The requirement tag is used to build traceability tables in order to ensure that all requirements were properly processed.

Predicates should not be too complex to be easily verified. To do so, definitions should be used intensively.
Actually, the production of definitions is partly done bottom-up, from interface constants to properties, and partly top-down, from properties to interface constants and partly by means of refactoring when we realise afterwards how to build a better model.

The following predicate corresponds to the requirement: \emph{position protected by a signal is located afterwards the signal regarding the signal direction}.

   \begin{center}
      \begin{minipage}{.8\textwidth}
	 \begin{description}
	       \Item{\forall mansig\qdot(mansig \in s\_mansig }
	       \begin{description} 
		     \Item{        \limp }
		     \Item{         f\_pos\_afterwards(f\_sig\_dir(mansig)      \mapsto  }
		     \begin{description}
			   \Item{                          f\_sig\_block(mansig)    \mapsto }
			   \Item{                          f\_sig\_abs(mansig)      \mapsto } 
			   \Item{                          f\_sig\_prot\_seg(mansig) \mapsto } 
			   \Item{                          f\_sig\_prot\_abs(sigman)   ) = \True) }
		     \end{description}
	       \end{description}
	 \end{description}
      \end{minipage}
   \end{center}

All given requirements have been modelled into predicates using the process described here. However the calculus on integers was not completely satisfactory, even though we worked with fixed point reals, integer computation introduces errors when using integer division. So in some cases, we had to replace an equality predicate, by a definition stating that two integers are distant from less than a given epsilon. Those epsilon values were explicitely put back into XML input files, so that the safety department could agree, or not, with those values. Fortunately, all the computations were pretty basic, so the limitations of the B language were not too much of a constraint.
However, if we had to model real floating point calculus with scientific operators like exponential or trigonometry operators, we would need to extend somehow the B language or the tools, but this is beyond the scope of this paper.

\subsection{Verifying and testing properties}

The B model is built property by property. After producing a property, the model developer exercises it against a nominal example through the tool chain based on Rodin plug-ins, since it is easier to use on unfinished models. If the property check fails, then the problem is tracked down just like software debugging. The Rodin plug-in tool displays the value (true/false) of all sub predicates, so we can explore the predicate, searching for something unexpected. This exploration is usually very effective and leads to the problem, which could be either a data, a model or a property error.

When the development of a property is complete, it should then be checked by a qualified person different from the person who formalised the property. This way, we make sure that the property is adequately formalised. However, experience shows that this process is not robust enough for complex properties. For these properties, we write one or more unit tests in order to test the different kinds of errors that could be expected from the predicate. Usually between one and three unit tests are written. Of course, this test strategy does not intend to be complete, but it is very efficient to point out small errors that could ruin the validation for that property. In one case, a definition providing a set of blocks was erroneously always equal to the empty set, implying that properties using the definition always succeeded.  

\subsection{Tool results}

The model was written with no particular concern about the tools, except for one rule: when defining a variable, instead of first typing it with pure typing predicates and then adding constraint predicates, it is far more efficient to directly type it with a constraint set expression.

In this data validation process, a double tool chain is used in order to mitigate the risk of error in these tools.

\begin{itemize}
\item The first tool chain is based on the Rodin AST plug-in and on the predicate evaluator core plug-in. The interface plug-ins are proprietary one.
\item The second tool chain is based on the ProB model checker. The interface part is also a proprietary one.
\end{itemize}

Both tools needed an adjustment of their evaluation strategy due to intensive use of closure and cartesian products with one argument being a large subset of integers. The tools had a tendency to perform an early evaluation of sets to be more efficient, however such a strategy does not pay off for closure or cartesian product evaluation for which a lazy strategy proves to be much more efficient.

After taking into account these two issues, both tools were very efficient. The validation of a set of configuration data containing several millons of integers took only a few minutes. No contradictory result was produced by the double chain which, given the high differentiation of the development tools ---both in process and implementation techniques--- gives a great confidence in the results.

\section{Pros and Cons of the Process}

\subsection{Pros}
A major benefit of this work is that the formal modelling activity focuses on the capital issue of understanding as clearly as possible the configuration data and their properties in all possible situations and not only for the simple situations one can think of or for the examples represented by diagrams of input documents.

One other major benefit of this work is that all properties, however hard to define on the railway graph, were entirely modelled in B. None was left to external error prone informal verification.
\begin{itemize}
   \item Well suited for CBTC configuration data: properties are neither too simple nor too complex.
   \item Reasonable validation time for both tools (a few minutes).
   \item The tools can handle large data (several megabytes).
   \item It is far more interesting to focus on high level formal modelling and on debugging data sets than to hard code verifications.
   \item It may also take less time to validate configuration data with predicate evaluation tools, than to develop safety-critical specific tools.
\end{itemize}

\subsection{Limitations}

\begin{itemize}
   \item Ill-suited for scientific calculus.
   \item From an industrial point of view, we are looking for other domains than railway safety-critical systems where this process can be applied with the same success.
\end{itemize}

\end{document}